\documentclass[aps,amsmath,amssymb,groupedaddress,twocolumn]{revtex4}
\usepackage{graphicx}

\newcommand{\nc}{\newcommand}
\nc{\beq}{\begin{equation}}
\nc{\eeq}{\end{equation}}
\nc{\beqa}{\begin{eqnarray}}
\nc{\eeqa}{\end{eqnarray}}

\def\gsim{\mathrel{\rlap{\lower4pt\hbox{\hskip1pt$\sim$}}
    \raise1pt\hbox{$>$}}}       

\begin{document}

\title{Information, information processing and gravity}

\author{Stephen~D.~H.~Hsu} \email{hsu@uoregon.edu}
\affiliation{Institute of Theoretical Science \\ University of Oregon,
Eugene, OR 97403}

\begin{abstract}
I discuss fundamental limits placed on information and information
processing by gravity. Such limits arise because both information and
its processing require energy, while gravitational collapse (formation
of a horizon or black hole) restricts the amount of energy allowed in
a finite region. Specifically, I use a criterion for gravitational
collapse called the hoop conjecture. Once the hoop conjecture is
assumed a number of results can be obtained directly: the existence of
a fundamental uncertainty in spatial distance of order the Planck
length, bounds on information (entropy) in a finite region, and a
bound on the rate of information processing in a finite region. In the
final section I discuss some cosmological issues, related to the total
amount of information in the universe, and note that almost all
detailed aspects of the late universe are determined by the randomness of quantum outcomes. This paper is based on a talk presented at a 2007 Bellairs
Research Institute (McGill University) workshop on black holes and
quantum information.
\end{abstract}


\maketitle

\date{today}

\bigskip

\section{Introduction}

This paper is based on a talk presented at a workshop on black holes
and quantum information (Bellairs Research Institute of McGill
University, Barbados).  Most of the
participants were quantum information theorists, so I attempted to keep the
technical details concerning general relativity or particle physics at
a minimum. I tried to summarize, in the most physical and intuitive way,
how gravity enforces some surprising constraints on information and
information processing. From a practical perspective, due to the
feebleness of the gravitational force, all of the limits deduced are
incredibly weak. Our technologies are nowhere near saturating them,
and they are of much greater interest to theoreticians than
experimentalists or engineers. Nevertheless, they are fundamental in
nature, and, depending as they do both on quantum mechanics and
general relativity, may offer a view into the properties of
quantum gravity.

In the discussion that follows, gravitational collapse will be our crude but
powerful probe of gravitational physics. Complete gravitational collapse 
leads to the formation of black holes and causal horizons. Gravity is 
a long range force that, as far as we know, {\it cannot be screened}. 
In this respect, it is fundamentally different
from gauge forces, such as the strong and electroweak interactions, 
and it is precisely this difference that allows for dramatic phenomena
like complete collapse. 

We use Planck units throughout, in which the speed of light, Planck's
constant and the Planck mass (equivalently, Newton's constant) are
unity. In our expressions, any energy or mass is therefore measured in
units of $10^{19}$ GeV (about $10^{-5}$ grams), and any length is
measured in units of the Planck length, or about $10^{-35}$ meters.

\section{Gravitational collapse}

Ideally, one would like to know precisely what subset of all possible
physical initial data results in gravitational collapse and the
formation of a black hole. This is obviously a difficult problem and
it is currently unsolved. Schoen and Yau \cite{SchYau} proved a
celebrated result requiring the existence of a closed trapped surface
if the {\it minimum} density in a region is sufficiently
high. However, this result fails to be useful if the energy of the
initial configuration is distributed in a very nonuniform manner.

Note that results for black hole or horizon formation typically
require both the assumption of the null or weak energy condition and
of cosmic censorship \cite{GR}. Under those assumptions a closed
trapped surface can be shown to result in a singularity (using the
Raychaudhuri equation and assuming the energy conditions hold), and
cosmic censorship requires a horizon to conceal the singularity from
asymptotic observers.

\begin{figure}
\includegraphics[width=5cm]{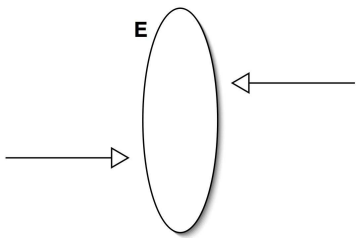}
\caption{The hoop conjecture applied to two relativistic particles.}
\label{hoop}
\end{figure}

In our analysis we will use the hoop conjecture, due to Kip Thorne
\cite{hoop} as a criterion for gravitational collapse. It states that
a system of total energy $E$, if confined to a sphere of radius $R <
\eta E$ ($\eta$ is a coefficient of order one, which we neglect
below), must inevitably evolve into a black hole. The condition $R <
E$ is readily motivated by the Schwarzschild radius $R_s = 2M$. This
conjecture has been confirmed in astrophysically-motivated numerical
simulations, and has been placed on even stronger footing by recent
results on black hole formation from relativistic particle collisions
\cite{bh}. These results show that, even in the case when all of the
energy $E$ is provided by the kinetic energy of two highly boosted
particles, a black hole forms whenever the particles pass within a
distance of order $E$ of each other (see Fig.~\ref{hoop}). Two
particle collisions had seemed the most likely to provide a
counterexample to the conjecture, since the considerable kinetic
energy of each particle might have allowed escape from gravitational
collapse.

One can think of the hoop conjecture as requiring that the average
energy density of an object of size $R$ be less than $R^{-2}$
in order not to collapse to a black hole. Thus, large objects which
are not black holes must be less and less dense. For example, a
sufficiently large object with only the density of water will
eventually form a black hole!

\section{Minimal length}

In this section we deduce a fundamental limit on our ability to
measure a distance \cite{CGH,minlength,clock,foam}.  The results
suggest that spacetime may ultimately have a discrete structure. At
the end of the section we discuss the implications for quantum
information and the ultimate Hilbert space of quantum mechanics.

From the hoop conjecture (HC) and the uncertainty principle, we
immediately deduce the existence of a {\it minimum ball} of size
$l_P$. Consider a particle of energy $E$ which is not already a black
hole. Its size $r$ must satisfy \beq r \gsim {\rm \bf max} \left[\,
1/E\, ,\,E \, \right]~~, \eeq where $\lambda_C \sim 1/E$ is its
Compton wavelength and $E$ arises from the hoop
conjecture. Minimization with respect to $E$ results in $r$ of order
unity in Planck units, or $r \sim l_P$ \cite{interval}.  If the
particle {\it is} a black hole, then its radius grows with mass: $r
\sim E \sim 1/ \lambda_C$. This relationship suggests that an
experiment designed (in the absence of gravity) to measure a short
distance $l << l_P$ will (in the presence of gravity) only be
sensitive to distances $1/l$. This is the classical counterpart to
T-duality in string theory \cite{duality}.

It is possible that quantum gravitational corrections modify 
the relation between $E$ and $R$ in the HC. However, 
if $E$ is much larger than the Planck mass, 
and $R$ much larger than $l_P$, we expect
semiclassical considerations to be reliable. (Indeed, in two particle
collisions with center of mass energy much larger than the Planck
mass the black holes produced are semiclassical.) This means that the
existence of a minimum ball of size much smaller than $l_P$ does {\it not}
depend on quantum gravity - the energy required to confine a particle
to a region of size much smaller than $l_P$ would produce a large, 
semiclassical black hole.

Before proceeding further, we give a concrete model of minimum length
that will be useful later. Let the position operator $\hat{x}$ have
discrete eigenvalues $\{ x_i \}$, with the separation between
eigenvalues either of order $l_P$ or smaller.  (For regularly
distributed eigenvalues with a constant separation, this would be
equivalent to a spatial lattice.)  We do not mean to imply that nature
implements minimum length in this particular fashion - most likely,
the physical mechanism is more complicated, and may involve, for
example, spacetime foam or strings. However, our concrete formulation
lends itself to detailed analysis. We show below that this formulation
cannot be excluded by any gedanken experiment, which is strong
evidence for the existence of a minimum length.

Quantization of position does not by itself imply quantization of
momentum. Conversely, a continuous spectrum of momentum does not
imply a continuous spectrum of position. In a formulation of
quantum mechanics on a regular spatial lattice, with spacing $a$
and size $L$, the momentum operator has eigenvalues which are
spaced by $1/L$. In the infinite volume limit the momentum
operator can have continuous eigenvalues even if the spatial
lattice spacing is kept fixed. This means that the displacement
operator \beq \label{disp} \hat{x} (t) - \hat{x} (0) = \hat{p}(0)
{t \over M} \eeq does not necessarily have discrete eigenvalues
(the right hand side of (\ref{disp}) assumes free evolution; we
use the Heisenberg picture throughout). Since the time evolution
operator is unitary the eigenvalues of $\hat{x}(t)$ are the same
as $\hat{x}(0)$. Importantly though, the spectrum of $\hat{x}(0)$
(or $\hat{x}(t)$) is completely unrelated to the spectrum of the
$\hat{p}(0)$, even though they are related by (\ref{disp})
\cite{dspec}.
Consequently, we stress that {\em a measurement of the
displacement is a measurement of the spectrum of $\hat{p}(0)$ (for
free evolution) and does not provide information on the spectrum
of $\hat{x}$.} A measurement of arbitrarily small displacement
(\ref{disp}) does not exclude our model of minimum length. To
exclude it, one would have to measure a position eigenvalue $x$
and a nearby eigenvalue $x'$, with $|x - x'| << l_P$.

Many minimum length arguments (involving, e.g., a microscope,
scattering experiment or even Wigner's clock \cite{minlength}) are
obviated by the simple observation of the minimum ball. However,
the existence of a minimum ball does not by itself preclude the
localization of a macroscopic object to very high precision.
Hence, one might attempt to measure the spectrum of $\hat{x}(0)$
through a time of flight experiment in which wavepackets of
primitive probes are bounced off of well-localised macroscopic
objects. Disregarding gravitational effects, the discrete spectrum
of $\hat{x}(0)$ is in principle obtainable this way. But,
detecting the discreteness of $\hat{x}(0)$ requires wavelengths
comparable to the eigenvalue spacing.  For eigenvalue spacing
comparable or smaller than $l_P$, gravitational effects cannot be
ignored, because the process produces minimal balls (black holes)
of size $l_P$ or larger. This suggests a direct measurement of the
position spectrum to accuracy better than $l_P$ is not possible.
%
The
failure here is due to
the use of probes with very short wavelength.

\begin{figure}
\includegraphics[width=4cm]{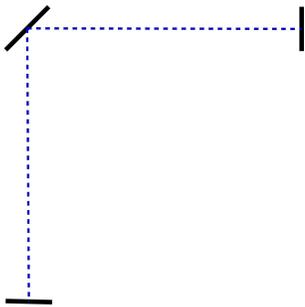}
\caption{An interferometer can be sensitive to path length differences
much smaller than the wavelength of light used.}
\label{interferometer}
\end{figure}

A different
class of instrument - the interferometer (Fig. \ref{interferometer}) 
- is capable of measuring
distances much smaller than the size of any of its sub-components
\cite{LIGO}. An interferometer can measure a distance \beq
\label{interferometer} \Delta x ~\sim~ { \lambda \over b \sqrt{N}}
~\sim~ {L \over \tau \sqrt{N} \nu}~~, \eeq where $\lambda = 1 /
\nu$ is the wavelength of light used, $L$ is the length of each
arm, $\tau$ the time duration of the measurement, and $N$
the number of photons. More precisely, $\Delta x$ is the change over
the duration of the measurement in the relative
path lengths of
the two arms of the interferometer.
$b = \tau / L$ is the number of bounces
over which the phase difference builds, so (\ref{interferometer})
can also be written as \beq \Delta \Phi ~=~ {b \Delta x \over
\lambda} ~\sim~ {1 \over \sqrt{N}}~~, \eeq which expresses
saturation of the quantum mechanical uncertainty relationship
between the phase and number operators of a coherent state.

From (\ref{interferometer}) it appears that $\Delta x$ can be made
arbitrarily small relative to $\lambda$ by, e.g., taking the
number of bounces to infinity. Were
this the case, we would have an experiment that, while still using
a wavelength $\lambda$ much larger than $l_P$,  could measure a
distance less than $l_P$ along one direction, albeit at the cost
of making the measured object (e.g., a gravity wave) large in the
time direction. This would contradict the existence of a {\it
minimum interval}, though not a minimum ball in spacetime.
(Another limit which increases the accuracy of the interferometer
is to take the number of photons $N$ to infinity, but this is more
directly constrained by gravitational collapse. Either limit is
ultimately bounded by the argument discussed below.)

A constraint which prevents an arbitrarily accurate measurement of
$\Delta x$ by an interferometer arises due to the Standard Quantum
Limit (SQL) and gravitational collapse. The SQL \cite{SQL} is
derived from the uncertainty principle (we give the derivation
below; it is not specific to interferometers, although see
\cite{noise}) and requires that \beq \label{SQL} \Delta x \geq
\sqrt{ t \over 2 M }~~, \eeq where $t$ is the time over
which the measurement occurs and $M$ the mass of the object whose
position is measured. In order to push $\Delta x$ below $l_P$, we
take $b$ and $t$ to be large. But from (\ref{SQL}) this
requires that $M$ be large as well. In order to avoid
gravitational collapse, the size $R$ of our measuring device must
also grow such that $R > M$. However, by causality $R$ cannot
exceed $t$. Any component of the device a distance greater than
$t$ away cannot affect the measurement, hence we should not
consider it part of the device. These considerations can be
summarized in the inequalities \beq \label{CGR} t > R > M
~~.\eeq Combined with the SQL (\ref{SQL}), they require $\Delta x
> 1$ in Planck units, or \beq \label{DLP} \Delta x > l_P~. \eeq
(Again, we neglect factors of order one.)

Notice that the considerations leading to (\ref{SQL}), (\ref{CGR})
and (\ref{DLP}) were in no way specific to an interferometer, and
hence are {\it device independent}. We repeat: no device subject
to the SQL, gravity and causality can exclude the quantization
of position on distances less than the Planck length.

It is important to emphasize that we are deducing a minimum length
which is parametrically of order $l_P$, but may be larger or
smaller by a numerical factor.  This point is relevant to the
question of whether an experimenter might be able to transmit the
result of the measurement before the formation of a closed trapped
surface, which prevents the escape of any signal. If we decrease
the minimum length by a numerical factor, the inequality
(\ref{SQL}) requires $M >> R$, so we force the experimenter to
work from deep inside an apparatus which has far exceeded the
criterion for gravitational collapse (i.e., it is much denser than
a black hole of the same size $R$ as the apparatus). For such an
apparatus a horizon will already exist before the measurement
begins. The radius of the horizon, which is of order $M$, is very
large compared to $R$, so that no signal can escape.

We now give the derivation of the Standard Quantum Limit. Consider
the Heisenberg operators for position $\hat{x} (t)$ and momentum
$\hat{p} (t)$ and recall the standard inequality \beq \label{UNC}
(\Delta A)^2 (\Delta B)^2 \geq  ~-{1 \over 4} ( \langle [
\hat{A}, \hat{B} ] \rangle )^2 ~~.
\eeq Suppose that the
position of a {\it free} test mass is measured at time $t=0$
 and {\em again} at a later time.
The
position operator at a later time $t$ is \beq \label{P} \hat{x}
(t) = \hat{x} (0) ~+~ \hat{p}(0) \frac{t}{M}~~. \eeq
The commutator between the position operators at $t=0$ and $t$
is \beq [ \hat{x} (0), \hat{x} (t)] ~=~ i {t \over M}~~,
\eeq so using (\ref{UNC}) we have \beq \vert \Delta x (0) \vert
\vert \Delta x(t) \vert \geq \frac{t}{2M}~~.\eeq
So, at least one of the uncertainties $\Delta x(0)$ or $\Delta x(t)$
must be larger than of order $\sqrt{t/M}$.
As a measurement of the discreteness of $\hat{x}(0)$
requires {\em two} position measurements,
it is limited by the greater of $\Delta x(0)$ or $\Delta x(t)$,
that is, by $\sqrt{t/M}$.

What are the consequences of a minimum length? In a discrete spacetime
there need not be any continuous degrees of freedom, and the number of
degrees of freedom in a fixed volume is finite. Further, one can show
that discretization of spacetime naturally suggests discretization of
Hilbert space itself \cite{Buniy:2005iw}. Specifically, in a universe
with a minimal length (for example, due to quantum gravity), no
experiment can exclude the possibility that Hilbert space is discrete.

\section{Entropy bounds}

In this section we describe two entropy bounds arising from
gravitational collapse. These bounds limit the number of degrees of
freedom in a region of size $R$, or equivalently the amount of
information in any system of fixed size.

The first bound uses the area--entropy relation for black holes.
Black holes radiate \cite{Hawking} and have entropy: $S = A / 4$
\cite{Bekenstein}. The nature of this entropy is one of the great
mysteries of modern physics, especially due to its non-extensive
nature: it scales as the area of the black hole (in Planck units),
rather than its volume. This peculiar property has led to the
holographic conjecture \cite{HoltHooft,HolSusskind} proposing that the
number of degrees of freedom in any region of our universe grows only
as the area of its boundary. (See \cite{Bousso} for a review and
discussion of covariant generalizations of this idea, and \cite{JO}
for a general discussion of how area bounds arise in gravitating
systems.) The AdS/CFT correspondence \cite{AdSCFTreview} is an
explicit realization of holography.

The entropy of a thermodynamic system is the logarithm of the number
of the available microstates of the system, subject to some
macroscopic constraints such as fixed total energy. In certain string
theory black holes, these states have been counted explicitly
\cite{SV, MSW}. 

\begin{figure}
\includegraphics[width=5cm]{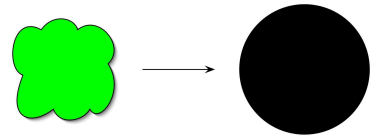}
\caption{A system collapses to form a black hole. By the second law of
thermodynamics, its original entropy is bounded above by that of the
black hole.}
\label{blob}
\end{figure}

Consider a system of size $R$ and total energy $E$ (e.g., the green
blob in Fig. \ref{blob}), which is not a black hole ($E < R$). Now
imagine a spherical shell of energy $R - E$ approaching the system at
the speed of light. By causality, the system is unaffected by the
shell until the combination of the two already satisfy the hoop
conjecture. The combined system must evolve into a black hole with
entropy $A / 4$. By the second law of thermodynamics, this final
entropy is larger than that of the initial system. Since we made no
particular assumptions about the initial system, we deduce that
ordinary (non-collapsed) physical systems have entropy less than their
surface area in Planck units. This is quite a counterintuitive result,
since in familiar (non-gravitating) systems entropy is typically
extensive.

The second bound, obtained by 't Hooft \cite{thooft}, shows that if one
excludes states from the Hilbert space whose energies are so large
that they would have already caused gravitational collapse, one
obtains $S = \ln N < A^{3/4}$, where $N$ is the number of degrees of
freedom and $A$ the surface area.  To deduce this result, 't Hooft
replaces the system under study with a thermal one. This is justified
because, in the large volume limit, the entropy of a system with
constant total energy $E$ (i.e., the logarithm of the phase space
volume of a microcanonical ensemble) is given to high accuracy by that
of a canonical ensemble whose temperature has been adjusted so that
the average energies of the two ensembles are the same. (This is a
standard, and central, result in statistical mechanics.)

Given a thermal region of radius $R$ and temperature $T$, we have $S
\sim T^3 R^3$ and $E \sim T^4 R^3$. Requiring $E < R$ then implies $T
\sim R^{-1/2}$ and $S < R^{3/2} \sim A^{3/4}$. We stress that the
thermal replacement is just a calculational trick: temperature plays
no role in the results, which can also be obtained by direct counting.

In \cite{Buniy:2005au}, it 
was shown that imposing the condition ${\rm Tr} [ \, \rho H \,] < R$
on a density matrix $\rho$ implies a similar bound $S_{\rm vN} <
A^{3/4}$ on the von Neumann entropy $S_{\rm vN} = {\rm Tr} \, \rho \ln
\rho$. For $\rho$ a pure state the result reduces to the previous
Hilbert space counting.  

We note that these bounds are more restrictive than the bound obtained
from black hole entropy: $S < A / 4$. One can interpret this
discrepancy as a consequence of higher entropy density of
gravitational degrees of freedom relative to ordinary matter
\cite{hsumurray}.

The consequences of these bounds are rather striking: they suggest
that gravitating systems in $d$ dimensions contain only as much
information as analogous, but non-gravitating, systems in $d-1$
dimensions. A concrete realization of this is the AdS/CFT duality in
string theory \cite{AdSCFTreview}.

\section{Bound on rate of information processing}

In this section we derive an upper bound on the rate at which a device
can process information \cite{compute}. We define this rate as the
number of logical operations per unit time, denoted as the ops rate
${\cal R}$. The operations in question can be those of either
classical or quantum computers. The basis of our result can be stated
very simply: information processing requires energy, and general
relativity limits the energy density of any object that does not
collapse to a black hole. Replacing {\it information processing} by
{\it information} in the previous sentence leads to holography or
black hole entropy bounds, a connection we will explore further
below. For related work on fundamental physical limits to computation,
see \cite{lloyd} and \cite{lloydng}.

Our result is easily deduced using the Margolus--Levitin (ML) theorem
\cite{lm} from quantum mechanics, and the hoop conjecture.

The Margolus--Levitin theorem states that a quantum system with
average energy $\epsilon$ requires at least $\Delta t > \epsilon^{-1}$
to evolve into an orthogonal (distinguishable) state. It is easy to
provide a heuristic justification of this result. For an energy
eigenstate of energy $E$, $E^{-1}$ is the time required for its phase
to change by order one. In a two state system the energy level
splitting $E$ is at most of order the average energy of the two
levels. Then, the usual energy-time uncertainty principle suggests
that the system cannot be made to undergo a controlled quantum jump on
timescales much less than $E^{-1}$, as this would introduce energy
larger than the splitting into the system.

Consider a device of size $R$ and volume $V \sim R^3$, comprised of
$n$ individual components \cite{fn1} of average energy
$\epsilon$. Then, the ML theorem gives an upper bound on the total
number of operations per unit time
\begin{equation}
{\cal R} < n \epsilon~,
\end{equation}
while the hoop conjecture requires $E \sim n \epsilon < R$. Combined, we obtain
\begin{equation}
\label{bound}
{\cal R} < R \sim V^{1/3}~.
\end{equation}

It is interesting to compare this bound to the rate of information
processing performed by nature in the evolution of physical
systems. At first glance, there appears to be a problem since one
typically assumes the number of degrees of freedom in a region is
proportional to $V$ (is extensive). Then, the amount of information
processing necessary to evolve such a system in time grows much faster
than our bound (\ref{bound}) as $V$ increases. Recall that for $n$
degrees of freedom (for simplicity, qubits), the dimension of Hilbert
space $H$ is $N = {\rm dim} \, H = 2^n$ and the entropy is $S = \ln N
\sim n$. In the extensive case, $n \sim S \sim V$.

However, as noted in the previous section, gravity also constrains the maximum
information content (entropy $S$) of a region of space. 't Hooft
\cite{thooft} showed that if one excludes states from the Hilbert
space whose energies are so large that they would have already caused
gravitational collapse, one obtains $S = \ln N < A^{3/4}$, where $N$
is the number of degrees of freedom and $A$ the surface area. 

Given this result we can calculate the maximum rate of information
processing necessary to simulate any physical system of volume $V$
which is not a black hole. The rate ${\cal R}$ is given by the number
of degrees of freedom $S \sim R^{3/2}$ times the maximal ML rate $T
\sim R^{-1/2}$. This yields ${\cal R} \sim R$ as in our bound
(\ref{bound}).

Finally, we note that black holes themselves appear to saturate our
bound. If we take the black hole entropy to be $S \sim A \sim R^2$,
and the typical energy of its modes to be the Hawking temperature
$T_{\rm H} \sim R^{-1}$, we again obtain ${\cal R} \sim R$.

\section{How much information in the universe?}

In this final section we ask how much information is necessary to
specify the current state of the universe, and where did it come from?

\begin{figure}
\includegraphics[width=6cm]{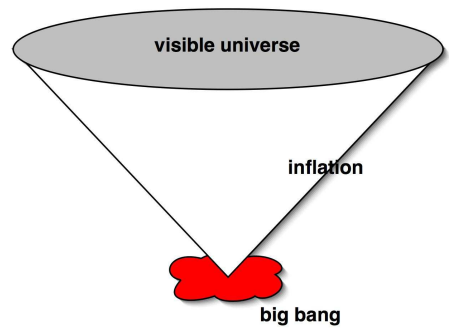}
\caption{The visible universe is exponentially larger than the initial
region from which it evolved.}
\label{inflation}
\end{figure}

There is convincing observational evidence for the big bang model of
cosmology, and specifically for the fact that the universe is and has
been expanding. In a radiation-dominated universe, the FRW scale
factor grows as $R(t) \sim t^{1/2}$, where $t$ is the comoving
cosmological time. From this, it is clear that our universe evolved
from a much smaller volume at early times. Indeed, in inflationary
cosmology (Fig.~\ref{inflation}) the visible universe results from an
initial patch which is exponentially smaller than our current horizon
volume. The corresponding ratio of entropies is similarly gigantic,
meaning that there is much more information in the universe today than
in the small primordial patch from which it originated. Therefore, the
set of possible early universe initial conditions is much, much
smaller than the set of possible late time universes. A mapping
between all the detailed rearrangements or modifications of the
universe today and the set of possible initial data is many to one,
not one to one \cite{rnd}.

Thus, the richness and variability of the universe we inhabit
cannot be attributed to the range of initial conditions. The fact that
I am typing this on a sunny day, or that our planet has a single moon,
or that the books on my office shelves have their current
arrangement, was not determined by big bang initial data.

How, then, do the richness and variability of our world arise? The
answer is quantum randomness -- the randomness inherent in
measurements of quantum outcomes.

Imagine an ensemble $\Psi$ of $n$ qubits, each prepared in an
identical state $\psi$. Now imagine that each qubit is measured, with
a resulting spin up ($+$) or spin down ($-$) result. There are $2^n$
possible records, or histories, of this measurement. This is an
exponentially large set of outcomes; among them are all possible
$n$-bit strings, including every $n$-bit work of literature it is
possible to write!  Although the initial state $\Psi$ contained very
little information (essentially, only a single qubit of information,
since each spin is in an identical state), $n$ bits of classical
information are required to specify {\it which} of the $2^n$ outcomes
is observed in a particular universe. For $n \rightarrow \infty$ the
set of possible records is arbitrarily rich and varied despite the
simplicity of initial state $\Psi$.

In the same way, given an initial quantum state $\Psi$ describing the
primordial patch of the big bang from which our horizon volume
evolved, one must still know the outcomes of a large number of quantum
measurements in order to specify the particulars of the universe
today. From a many worlds perspective, one must specify all the
decoherent outcomes to indicate a particular branch of the
wavefunction -- a staggering amount of information. Equivalently, from
the traditional Copenhagen perspective, each quantum measurement
injects a bit (or more) of truly random information into our universe,
and this randomness accounts for its variability.

The most familiar cosmological quantum randomness comes from
fluctuations of the inflaton field, which determine the spectrum of
primordial energy density fluctuations. It is these density
fluctuations that determine the locations of galaxies, stars and
planets today. However, from entropic or information theoretic
considerations we readily deduce that essentially {\it every} detailed
aspect of our universe (beyond the fundamental Lagrangian and some
general features of our spacetime and its contents) is a consequence
of quantum fluctuations!

\begin{figure}
\includegraphics[width=4cm]{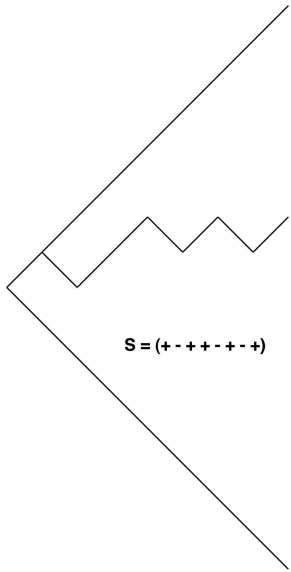}
\caption{Each path represents a set of outcomes $S$ from the
measurement of $n$ qubits. To specify a particular path requires
$n$ bits of classical information.}
\label{tree}
\end{figure}

\bigskip
\bigskip

\emph{Acknowledgements---} The author thanks the participants of the
Bellairs workshop, and especially the organizer Patrick Hayden, for a
stimulating and pleasant environment. A. Zee provided some important
comments on an earlier draft of the manuscript. The author is
supported by the Department of Energy under DE-FG02-96ER40969.



\bigskip



\begin{thebibliography}{99}
\bibitem{interval}
We focus specifically on spacelike minimum intervals. By
Lorentz covariance, we might expect that no spacetime interval 
$ds^2$ can be
measured to accuracy better than $l_P^2$, but this is a more subtle
issue.
\bibitem{dspec} In general the discrete spectrum of an observable $A(0)$
is unrelated to that of $A(t)-A(0)$. For instance, consider a
Hamiltonian $H=\alpha \sigma_z$ and $A(0)=\sigma_x$. Then the
spectrum of $A(0)$ is $\pm1/2$, but the spectrum of
$\sigma_x(t)-\sigma_x(0)$ is $\pm  2 \sin (\alpha t/2)$.
\bibitem{fn1} We identify $n$ individual components for later
comparison with entropy. The ML theorem could, of course, be applied
to the entire device as well.
\bibitem{rnd} Of course, it is possible that not all seemingly innocuous rearrangements of our universe are possible. Perhaps only an infinitesimal subset are possible future evolutions of the initial data -- this would be the case in a purely classical or deterministic universe. The situation would be analogous to a pseudo-random number generator: from a finite seed of limited entropy, it produces a much larger, seemingly random and aperiodic sequence of output. However, varying over all possible seeds does not produce all possible output strings, since the process is deterministic and the seed entropy is small.
\end{thebibliography}
\end{document}